# ON THE BIMODAL NATURE OF THE DISTRIBUTIONS OF PULSAR GLITCH SIZES

I. O. Eya[1,3,*], E. U. Iyida[2,3], C. I. Eze[2,3]

[1]Department of Science Laboratory Technology, University of Nigeria, Nsukka
[2]Department of Physics and Astronomy, University of Nigeria, Nsukka
[3]Astronomy and Astrophysics Research Lab, Faculty of Physical Sciences, University of Nigeria, Nsukka
*Corresponding Author: innocent.eya@unn.edu.ng

**Abstract**

Pulsar glitches are sudden spin-ups in pulsar spin frequency ($\nu$). The glitch size, $\Delta\nu/\nu$, is a key parameter in any mechanism puts across towards the understanding of the events. However, the distribution of the glitch sizes has persistently been bimodal. This bimodal nature could be intrinsic or otherwise. In this paper, the glitch size distribution is bisected at $\Delta\nu/\nu = 10^{-7}$ with $\Delta\nu/\nu < 10^{-7}$ being regarded as small size glitches (SSG), while those of $\Delta\nu/\nu > 10^{-7}$ are taken as large size glitches (LSG). The magnitude of SSG is scaled to that of LSG and tested for similarity. In pulsars with mixture of glitch sizes, Lilliefore test is used to identify the distribution pattern of SSGs and LSGs in such pulsars. The result indicates that each half of the size distribution is fundamentally different from one another. LSGs are seen to be normally distributed, while the SSGs are lognormal.

## 1. Introduction

The core remnants of supernova events, which manifest as pulsars are the most stable natural rotators in the present universe. This arguably is attributed to their huge moment of inertia ($\approx 10^{45}$ gcm$^2$). Nonetheless, long term Pulsar Timing have shown some form of rotational instability in pulsar timing data (e.g. Radhakrishnanand Manchester, 1969; Boynton et al., 1969; Helfand et al., 1980; Cordesand Downs, 1985; ChukwudeandUrama, 2010; Espinoza et al., 2011). The most spectacular one among the rotational instabilities is the pulsar glitch. Conventionally, pulsar glitches are seen as a sudden increase in pulsar spin frequency, $\Delta\nu$, which is sometimes accompanied by a change in frequency derivative, $\Delta\dot{\nu}$, (Espinoza et al.,2011; Yu et al., 2013). Most of the theories propounded towards the understanding of glitch events in pulsars are centred on the glitch sizes and the frequency of the events.

Just after the first observed pulsar glitch, starquake was suspected to be the cause of the event (Ruderman, 1969). Starquake model is based on the fact that a sudden reduction in moment of inertia, would lead to a sudden increase in angular frequency. Furthermore, it is presumed that due to high spin rate of pulsars, they are more oblate in shape than spherical. As such, they are off dynamic equilibrium and series of quake to reduce the oblateness to attain stable equilibrium is inevitable.Each of the quakes involves a release of an elastic energy accumulated in spin down process.The elastic energy is proportional to the glitch size, and the glitch size is proportional to the reduction in the oblateness (Baymand Pines, 1971; Zhou et al., 2014). These proportionalities constrain the frequency of glitch events in a given pulsar.In this framework, a reduction in oblateness of the order of centimetre, readily trigger a glitch size of $\Delta v/v \approx 10^{-9}$.This glitch size is in line with Crab pulsar glitches.However, Vela pulsar glitches ($\Delta v/v \approx 10^{-6}$) could not be well understood in the starquake model. This is because in a starquake model, Vela-like pulsars are expected to release an energy $\Delta v/v \approx 4\times 10^{36}$ erg per glitch (Baym& Pines, 1971; Zhou et al., 2014)} in a short time of less than a minute (Dodson et al., 2007).When this amount of energy is compared to the rotational power of Vela pulsar ($6\times 10^{36}$ erg s$^{-1}$), it indicates that Vela pulsar will not be able to glitch more than once in a lifetime.Yet Vela pulsar is seen to glitch every two to three years.These controversies lead to an alternative model – the angular momentum transfer model.

The angular momentum transfer model, though of many versions (e.g. Baym et al., 1969; Anderson & Itoh, 1975; Alpar et al., 1984b; Link & Epstein, 1996; Carter et al., 2000) relies mainly on transfer of angular moment from pulsar interior to the outer-crust. In it, it is assumed that the inner-crust of a pulsar contains a superfluid, which rotates differentially via an array of quantized vortices. The vortices are pinned to the lattice of the inner-crust leading to partial decoupling of the superfluid from the rest of the star.The velocity of the superfluid is proportional to the number of vortices it possesses and it is higher than that of the outer-crust.At a right condition that is not well understood, the vortices unpin and transfer their momentum to the outer-crust. This leads to sudden spin-up in pulsars known as glitches.Though the angular momentum transfer model accommodates the frequency of Vela-like glitches, the theoretical magnitude of the inner-crust is not in line with recent observational magnitude of the inner-crust as constrained by glitch sizes (Andersson et al., 2012; Chamel, 2013; Eya et al., 2017; Basu et al., 2018; Eya et al., 2019a). Thus, the cause of pulsar glitches remains an open debate.

The distribution of fractional glitch sizes, $\Delta v/v$, is in the range of $10^{-11}$ - $10^{-5}$ (Espinoza et al., 2011), while that of the inter-glitch time intervals, $\Delta t$, is in the range of 20 d - $10^4$d (Eya et al., 2019b).Recently, the distribution of fractional glitch size was successfully bisected at $\Delta v/v = 10^{-7}$ with the distribution of each half being similar when divided by their corresponding inter-glitch time interval (Eya et al., 2019b).In the analysis, glitches of $\Delta v/v < 10^{-7}$ are regarded as small size glitches (SSG), while those of $\Delta v/v > 10^{-7}$ are taken to be relatively large size glitches (LSG).Actually, $\Delta v/v = 10^{-7}$ corresponds to the reoccurring dip in the distribution of fractional glitch sizes, which made the distribution to be consistently bimodal (Wang et al., 2000; Espinoza et al., 2011; Yu et al., 2013; EyaandUrama, 2014; Eya et al., 2017;Eze et al., 2018; Eya et al., 2019b).

Meanwhile, Vela pulsar (PSR J0835-4510) and Crab pulsar (J0534+2200) were the first to record glitch event (Radhakrishnanand Manchester, 1969; Boynton et al., 1969).Interestingly, the characteristic size of their glitches is at the opposite ends of the glitches size distribution.Vela pulsar glitches are mainly of the order of $\Delta v/v \approx 10^{-6}$ and that of Crab pulsar are of the order of $\Delta v/v \approx 10^{-9}$.These two values correspond to the peaks of the bimodal distributions of pulsar glitch sizes.As such the lower end of the distribution is most of the time refereed to as Crab-like glitches, while the upper end is refereed to Vela like glitches, However it was recently reported that Crab pulsar had a large glitch comparable to that of Vela pulsar (Shaw et al., 2018) and Vela pulsar also had a glitch that is smaller than those of Crab pulsar (Jankowski et al., 2015). With these reports and recent analysis that showed that the inter-glitch time intervals are size independent (Eya et al., 2019a), it become imperative to study the bimodal nature of the glitch sizes distribution to ascertain if the dual nature is intrinsic or extrinsic. It is intrinsic in the sense that it is just by chance and has nothing to do with the idea of dual glitch mechanisms and extrinsic that it has much to do with glitch mechanisms. In this paper, each half of the distribution (as bisected by $\Delta\nu/\nu = 10^{-7}$) is analysed statistically to ascertain the nature of the bimodal distribution.

## 2. Data Analyses and Results

The data for this analysis is from the Jodrell Bank Observatory (JBO) glitch catalogue. The catalogue contains 483 glitches in rotation of 168 pulsars as at the time of this analysis (for update one can see http://www.jb.man.ac.uk/pulsar/glitches.html). The fractional sizes of the glitches in the catalogue are in the range $10^{-12} < \Delta\nu/\nu < 10^{-4}$, while the related change in the spin-down rate is of the order of $10^{-6} \leq |\Delta\dot{\nu}/\dot{\nu}| \leq 1$. The distributions of the fractional glitch sizes, $\Delta\nu/\nu$, is shown in Figure 1. The distribution is continuous and bimodal as usual (Wang et al., 2000; Espinoza et al., 2011; Yu et al., 2013; Eya and Urama, 2014; Eya et al., 2017, 2019b) with peaks at $\Delta\nu/\nu \sim 2.1 \times 10^{-9}$ and $\Delta\nu/\nu \sim 1.2 \times 10^{-6}$. The usual dip that demarcates SSG from LSG is at $\sim 10^{-7}$. The measures of the central tendency of the distributions as bisected by the dip are: mode $= 1.76 \times 10^{-6}$, mean $= 1.77 \times 10^{-6}$, median $= 1.76 \times 10^{-6}$ for LSGs and mode $= 2.14 \times 10^{-9}$, mean $= 2.88 \times 10^{-9}$, median $= 2.82 \times 10^{-9}$ for SSGs respectively.

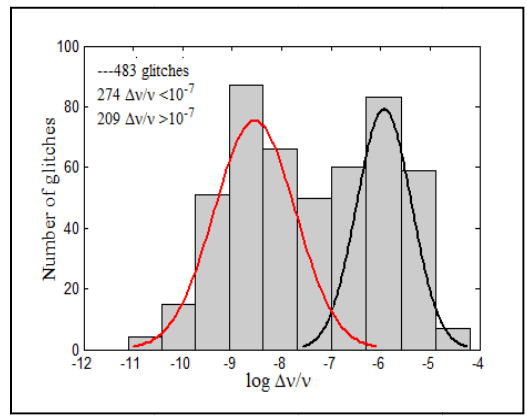

*Fig.1 Distribution of fractional glitch sizes ($\Delta\nu/\nu$). (**Note**: for more information on bisecting of glitch size distribution at $Log\Delta\nu/\nu = 10^{-7}$, see Eya et al., (2019a))*

In order to investigate whether the distributions of the two classes of glitch sizes (i.e. SSG and LSG) are the same, SSGs are scaled by a factor 549.54 (this is the ratio of LGS modal peak to SSG modal peak), so that its modal peak coincides with that of LSG, and explore the two dimensional Kolmogorov-Smirnov (K-S) test. The scaling is important so that the mode of the two distributions will have the same value. The cumulative distribution function (CDF) plot of the data is used to present the result pictorially. In the test, if two datasets, say, $x_1$ and $x_2$ illustrated by CDFs $F_1(x)$ and $F_2(x)$ are drawn from a common continuous distribution x, the null hypothesis is true (i.e. h = 0). In alternative, if the the datasets are drawn from different kinds of distribution, the null hypothesis is false (i.e. h = 1). The K-S Statistic K is the maximum displacement of one of the CDF from the other. That the result obtained is by chance is ascertained from P-value (Note: the values of K and P are in the range of 0 – 1). The null hypothesis is not reliable for a small P-value, however, a small P-value is a strong evidence for accepting the alternative hypothesis (i.e. h = 1). The P-value becomes accurate when ratio $(n_1 n_2)/(n_1 + n_2) \geq 4$, where $n_1$ and $n_2$ are the number of element in $F_1(x)$ and $F_2(x)$ respectively. In this analysis, ratio is equal to 118.56 enabling us to trust the P-value. In addition, the precision of the test depends on the confidence level of the test. In this paper, the K-S tests were performed at 95% confidence level. If the null hypothesis is true for the distributions of SSG and LSG, it follows that similar mechanisms are responsible for both classes of glitches, else otherwise.

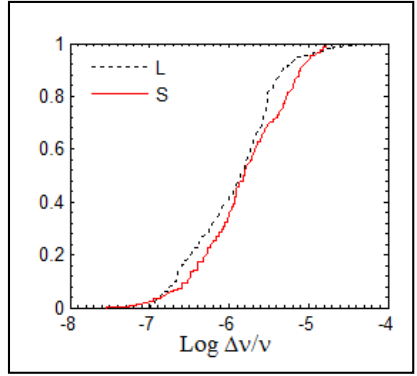

*Fig 2: CDF plots of classes of glitch sizes of which their modal peaks are coincident. L denotes LSG ($\Delta v/v > 10^{-7}$), while S denotes SSG ($\Delta v/v < 10^{-7}$).*

Table 1: Some K-S test results on the distributions of LSG and SSG scaled by α (Note: in all, h = 1).

| A | P | K |
|---|---|---|
| 200 | $2.42 \times 10^{-5}$ | 0.221 |
| 220 | $1.97 \times 10^{-4}$ | 0.202 |
| 240 | $4.79 \times 10^{-4}$ | 0.190 |
| 260 | $2.40 \times 10^{-3}$ | 0.174 |
| 280 | $4.80 \times 10^{-3}$ | 0.163 |
| 300 | $1.00 \times 10^{-3}$ | 0.150 |
| 320 | $8.70 \times 10^{-2}$ | 0.153 |
| 340 | $3.50 \times 10^{-3}$ | 0.164 |
| 360 | $1.90 \times 10^{-3}$ | 0.172 |
| 380 | $1.10 \times 10^{-3}$ | 0.180 |

The results obtained from the application of the K-S test to the two classes of the glitch sizes is h = 1, K = 0.22, and P = 1.44 ×10$^{-5}$. The CDFs is presented in Figure 2. The outcome of this preliminary test indicates that the distribution of SSG is significantly different from that of LSG. On the other hand, as some of the the elements in the distributions of both classes of glitches are outside the Gaussian fits, the distribution of SSG and LSG could also be the same at other measure of central tendencies other than the modes. To investigate this, SSGs are scaled in such a way that the mean/median of SSG coincides with that of LGS using a factor of $α_{mean}/α_{median}$= 407.38/416.87 and repeated the K-S test. The results also showed that the two distributions are significantly different. The P-values are $5.67 \times 10^{-4}$ and $3.78 \times 10^{-4}$ for $α_{mean}$ and $α_{median}$ respectively with K-value of 0.182 and 0.186. The CDFs are shown in Figure 3 (topmost panel). Meanwhile from the values of the mean and the median of SSG, it indicates that there are slightly more data at the left-hand-side of the mean (negative skewness). Therefore, to investigate weather there is a point below α$_{mean}$where the two distributions could be the same;the SSG is scaled by a factor of $200 \leq \alpha < \alpha_{mean}$ in steps of 20.The K-S test was repeated at each interval. The results are shown in Table 1 and the corresponding CDF plots shown in Figure 3.The outcome of this third test also indicates that the distributions of SSG and LSG are not the same. The closest for the two distributions to look alike is when $\alpha = 300$, beyond which, the magnitude of K increases progressively while P-value gets smaller, suggesting that the two samples are increasingly different. The P-value at the interval (i.e. $\alpha = 300$) is not significantly large to suggest that the null hypothesis is rejected by chance. With this, it suggests that the distribution of SSG is intrinsically different from that of LSG.The conditions responsible for regulating SSG are presumably different from that of LSG.

Whether a pulsar exhibit a specific glitch size could be studied with CDF curve (Espinoza et al., 2011). The CDF of glitch sizes should have a trend, with narrow dispersion about a mean value in such a pulsar (Eya et al., 2017). In order to investigate how each class of glitch sizes is distributed in pulsars with a mixture of glitch sizes, their CDFs wereanalysed. In this stage of the analysis, effort is concentrated on pulsars that have at least seven glitches. In such pulsars, six of the glitches must belong to either LSG or SSG (this is the minimum number of glitches that will be used in the analysis). Pulsars in which the number of glitches they have met these criteria are ten in number and the results of the analyses are summarized in Table 2. Interestingly, the number of glitches in three of the pulsars met the criteria for both classes, namely, PSRs J1341-6220, J1740-3015, and J1801-2304. In what follows, the CDF of each class of glitch sizes in the selected pulsars is examined to ascertain the type of distribution it mimics. To do this, each of the CDF is examined with Lilliefore'stest at 95% confidence level (note: Lilliefore test determines whether in a given dataset, if the elements in the dataset are similarly sized and symmetrical about a mean value). The test could identify class of glitches in which the glitches originated from momentum reservoir of similar vortex configuration, as glitch size depends on the number of vortices involved, the location of the vortices and the distance the vortices travelled before repining(Warszawski and Melatos, 2011). If approximately equal number of vortices were unpinned at each event in a given class of glitch sizes, and they migrate similar distance before repining, the distribution should be normal. The result of the test is presented with CDF plot and the CDF of ideal distribution it mimics is fitted on it. Unlike the one dimensional K-S test, the Lilliefors test does not require to predetermine the type of normal distribution the CDF mimic, instead the type of normal distribution is determined from the dataset (Lilliefors, 1967: 1969). A null hypothesis in Lilliefors test (i.e. h = 0), is that the elements in the dataset are drawn from a normal distributed population or else, otherwise. The K-value in Lilliefors test is the maximum difference between the empirical distribution estimated from the sample and a distribution with mean and standard deviation equal to the mean and standard deviation of the sample.

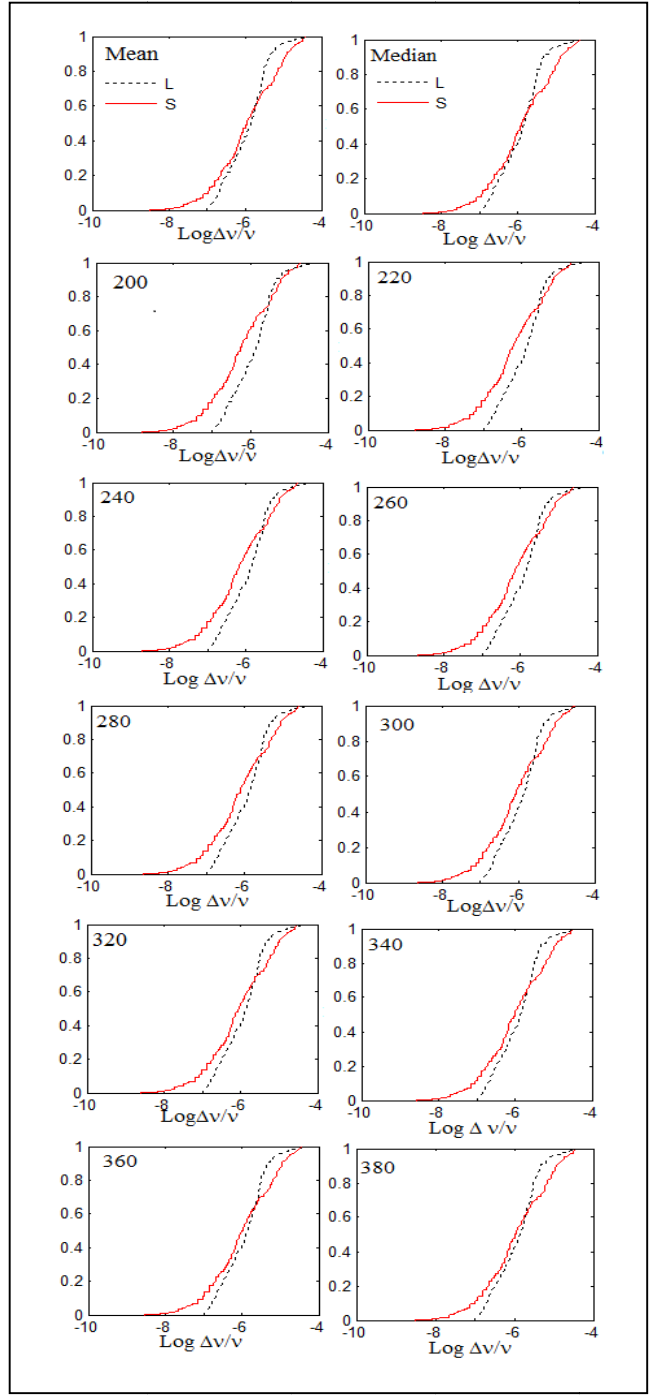

*Fig 3: Distribution of $\Delta v/v < 10^{-7}$ scaled by a factor α. Topmost panel are the mean and median of SSG scaled to the corresponding values of LSG. The numbers in other plots indicate the magnitude the SSGs were scaled. The solid curve indicates SSGs scaled by α.*

On application of Lilliefors test on the two classes of glitches, the results indicate that the distribution of LSG in the pulsars is normal except in pulsar J1708 − 4009 (which has h = 1, K = 0.02 and P = 0.36). However, for the SSGs, the null hypothesis is false, but on transforming the SSGs to logarithmic scale, the distribution became normal: implying that the distribution of SSGs in the pulsars is lognormal. The results are summarised in Table 3, while the CDF plots are shown in Figures 4 and 5 for LSGs and SSGs respectively. The fitson the CDF plots indicate how the glitch data fit the ideal distribution they mimic. To quantify numerically the goodness of the fits, the two dimensional K-S test is explored. The test indicates that the null hypothesis is true. The corresponding P-values are shown in Table 3. The outcome of this test suggests that the conditions necessary for LSG is the same in the pulsars sampled except in pulsar J1708 – 4009. A similar result has been obtained earlier using the entire glitch spin-up sizes $\Delta \nu/\nu$ in a PSRs J0537-6910, J0835-4510 and J1341-6220 (Eya et al., 2017). This result suggests that, approximately equal numbers of vortices were involved in the glitch process of LSGs (assuming angular momentum transfer process is solely responsible for the glitches). On the side of SSG, a lognormal distribution was reported in Crab pulsar glitches using the entire glitches (Espinoza et al., 2014). Though in the analysis, it is argued that power law distribution suits the distribution better. This could be as a result of mixture of both classes of glitch sizes in their analysis. The lognormal distribution of SSG indicates that multiple mutually independent parameters might be involved in regulating the glitch sizes. Such parameters includes the number of vortices involved, the distance unpinned vortices could migrate before repining, the temperature of the vortex pinning region, or the ellipticity of the star prior to the glitch assuming star-quake model. However, these conclusions are still tentative due to the small number of glitches involved. On the other hand, from the magnitude of K-values, the null hypothesis is fairly reliable as it indicates a small difference between the ideal distribution the sample mimics and that of the sample.

Table 2: Pulsar with mixture of glitches sizes

| Pulsar name | $N_g$ | $N_{>-7}$ | Pulsar name | $N_g$ | $N_{<-7}$ |
|---|---|---|---|---|---|
| J0537-6910 | 23 | 21 | J0534+2200 | 25 | 24 |
| J0835-4510 | 19 | 17 | J0631+1036 | 15 | 13 |
| J1341-6220 | 23 | 16 | J0742-2822 | 8 | 8 |
| J1708-4009 | 6 | 6 | J1341-6220 | 23 | 7 |
| J1740-3015 | 35 | 10 | J1740-3015 | 35 | 25 |
| J1801-2304 | 13 | 6 | J1801-2304 | 13 | 7 |
| | | | J1814-1744 | 7 | 7 |

Table 3: Summary of Lilliefors test (Note: h = 0 in all).

| LGS | K | P | SGS | K | P |
|---|---|---|---|---|---|
| J0537-6910 | 0.11 | 0.63 | J0534+2200 | 0.07 | 0.94 |
| J0835-4510 | 0.14 | 0.60 | J0631+1036 | 0.17 | 0.86 |
| J1341-6220 | 0.17 | 0.56 | J0742-2822 | 0.13 | 0.82 |
| J1740-3015 | 0.12 | 0.57 | J1341-6220 | 0.18 | 0.61 |
| J1801-2304 | 0.15 | 0.68 | J1740-3015 | 0.13 | 0.91 |
| | | | J1801-2304 | 0.12 | 0.66 |
| | | | J1814-1744 | 0.06 | 0.71 |

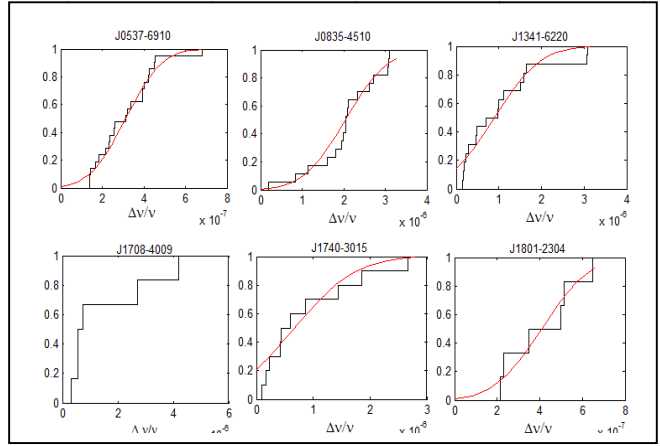

*Fig 4:CDF plots of LSG in individual pulsar with a normal fit, which has the same mean and standard deviation as the data. There is no fit on CDF curve of PSR J1708 - 4009 because the result of the test is not inaffirmative.*

### 3. Discussion

The dip in the distribution of glitch sizes (Figure 1) has remained one of the strongest evidence in favour of dual glitch mechanisms (Yu et al., 2013).This notion is revalidated by the discordant noticed in the distribution of the two classes of glitch sizes even when the central tendencies of the two distributions are coincident.Furthermore, the result of the lilliefore test in the distribution of glitch sizes not being the same in both classes of glitches is also a support to the argument in favour of dual glitch mechanism.This is because the elements in each class are similarly sized with respect to the distribution pattern they mimic.This is an indication that they could be of a common origin; the conditions that culminate in glitch sizes in a given class are similar. In such a scenario, the elements in a given class are mutually independent of that from the other class.

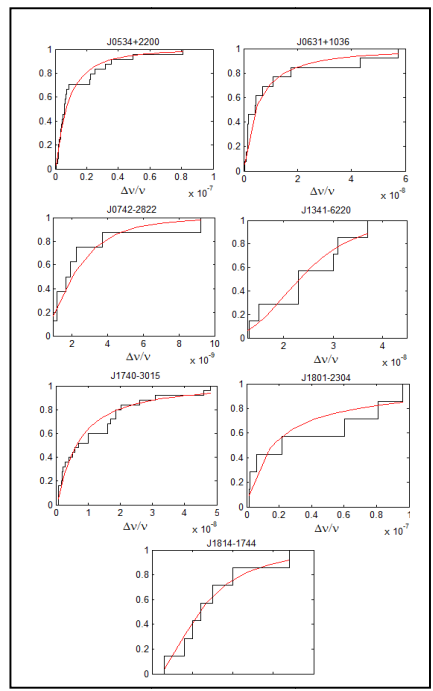

*Fig 5:CDF plots of SSGs in individual pulsar with a lognormal fit, which has the same mean and standard deviation as the data.*

Meanwhile, a recent analysis shows that angular momentum transfer model could readily account for current range of glitch sizes, including the missing ones that manifests in a reoccurring dip in the distribution of glitch sizes (Eya et al., 2017).The normality in the distribution of LSG in individual pulsars is a consequence of approximately equal number of unpinned vortices travelling similar distances before repining.This observation suggests that LSGs in a given pulsar originate from superfluid layer of similar vortex configuration.

For PSR J1708 – 4009, the abnormality in the distribution of LSG in it could be due to generic factor.This object is an anomalous X-ray pulsar (AXP) of which there are evidences of magnetospheric activity during AXPs glitches (Dib et al., 2008; KaspiandBeloborodov, 2017), which might have affected their glitch sizes. In that frame, what is measured may not be from angular momentum transfer alone and may lead to a wide dispersion in the distribution of its glitch sizes. For the SSGs, the distribution not being normal but lognormal is an indication that the factors that cumulated to the glitch sizes are not similar in each other.It could be that unequal numbers of vortices were unpinned or that the unpinned vortices do not travel similar distances before repining.

However, Figure 3 and Table 1 show that the distributions of the two classes of glitches could not be made the same by scaling the magnitude of one to the other. Apparently, this is an indication that the difference between SSG and LSG is not in magnitude via the number of vortices involved alone, but could relate to some other unclear processes in the interior of neutron star. There is a possibility that regions of distinct pinning strength exist within the inner-crust of neutron star (Alpar et al., 1984a). In these regions, the vortex motion could be linear or non-linear depending on the stellar temperature (Alpar et al., 1989). As such, in cooler pulsars with homogeneous pinning regions, the vortex motion could be linear leading to glitches of similar size, which are normally distributed.  For hot pulsars, the vortex pinning regions could be heterogeneous and vortex motion non-linear leading to varying glitch size. Glitches from linear region could be those whose sizes are normally distributed, while those, which are non-linear, are log-normally distributed.

In conclusion, the reoccurring dip in the distribution of glitch size is not by chance; instead it has much to do with glitch mechanism, which is still not well understood.